\documentclass[twocolumn,showpacs,amsmath,floatfix]{revtex4}
\usepackage[utf8]{inputenc}
\usepackage{amsmath,amssymb,amsfonts}
\usepackage[T1]{fontenc}
\usepackage{verbatim}
\usepackage{textcomp}
\usepackage{amsmath}
\usepackage[toc,page]{appendix}
\usepackage{tabularx}
\usepackage{float}
\usepackage[caption = false]{subfig}
\usepackage{graphicx}
\usepackage{lscape}
\usepackage{float}
\usepackage{xcolor}
\usepackage{longtable}
\usepackage{lmodern}
\usepackage{multirow}
\usepackage{booktabs, cellspace}
\definecolor{mygray}{gray}{0.3}

\begin{document}
\title{Ferroelectricity in oxyfluoride Aurivillius phase Bi$_2$TiO$_4$F$_2$ from first-principles}
\author{Sarah Benomar}
\affiliation{Laboratoire de Chimie Th\'eorique Computationnelle et Photonique, facult\'e de Chimie, Universit\'e Houari Boumedienne, BP.32, El Alia, Bab Ezzouar, Alger, Algeria}
\author{Hania Djani}
\affiliation{Centre de D\'eveloppement des Technologies Avanc\'ees, Cit\'e 20 ao\^ut 1956, Baba Hassen, Alger, Algeria}

\begin{abstract}
Using first-principles calculations, we restrained the m=1 oxyfluoride Aurivillius Bi$_2$TiO$_4$F$_2$ to adopt three ordered key configurations, such as tetragonal $I4/mmm$ phases with fluorine anions distributed either at the octahedral apical sites or at the octahedral equatorial ones and orthorhombic $Pma2$ phase with fluorine anions distributed over both apical and equatorial sites. We explored the energetics of the metastable phases and their potential for ferroelectricity and found that the least favorable configuration is equatorial sites for fluorine and the most favorable one is the mixture between apical and equatorial sites for fluorine. 
Phonon calculations performed in the tetragonal $I4/mmm$ with fluorine at the apical sites showed a strong c-axis polar  instability $\Gamma_{3}^{-}$  that co-exists with the ab-plane polar instability $\Gamma_{5}^{-}$, commonly found in Aurivillius oxide phases. These two polar instabilities lead to $Pc$ phase that displays in-plane and out-of-plane spontaneous polarizations as large as 44 $\mu$c/cm$^{2}$ and 35 $\mu$c/cm$^{2}$, respectively, opening new prospective design in thin films Aurivillius FeRAM devices.

\end{abstract}
\maketitle

\section{introduction}


Aurivillius phase structures family, with the general formula Bi$_2$A$_{m-1}$B$_m$O$_{3m+3}$ are mostly ferroelectric oxides, consisting of alternating, along the c-axis, of [Bi$_2$O$_2$]$^{+2}$ layers and perovskite-like (A$_{m-1}$B$_m$O$_{3m+1}$)$^{-2}$ blocks, with, A= Ca$^{2+}$, Ba$^{2+}$, Sr$^{2+}$, Bi$^{2+}$ ..., B= Fe$^{3+}$, Ti$^{4+}$, Nb$^{5+}$, W$^{6+}$ .., and $m$ is the number of BO$_6$ octahedra in the perovskite-like block ($m$=1-8) and is interdependent with A and B cations oxidation states to always provide right number of electrons to oxygen atoms~\cite{aurivillius49, perez-mato08, withers91, perez-mato04, etxebarria, hervoches}.
Typically, oxide Aurivillius phase structures crystallize at high temperature into centrosymmetric $I4/mmm$ tetragonal phase and adopt at room temperature a polar orthorhombic ground state involving in-plane polar displacement and in-plane and out-of-plane octahedral tilts about [110]$_t$ and [001]$_t$ directions, respectively ($t$ referring to the tetragonal parent structure). They are also known to exhibit a large spontaneous polarization P$_s$ $\approx$ 30-50 $\mu$c/cm$^{2}$ and high Curie temperature (T$_c$ > 350°C). Bi$_2$W0$_6$ is the smallest member with $m=1$. It is a one-layer strong  ferroelectric with  A cation deficiency in the perovskite block, with P$_s$ = 50 $\mu$c/cm$^{2}$ and T$_c$ = 950°C.

Oxyfluoride Aurivillius phases are also naturally occuring. The $m$=1  Bi$_2$TiO$_4$F$_2$, isostructural to  Bi$_2$NbO$_5$F and Bi$_2$TaO$_5$F, was first reported in 1952 ~\cite{aurivillius53} and has been intensively studied because of its ferroelectric and photocatalysis properties~\cite{C5RA14288A, C1DT10889A}. The first experimental study  claimed that Bi$_2$TiO$_4$F$_2$ was ferroelectric with a Curie temperature  T$_c$ $\approx$ 12°C and that the ferroelectric phase transition was accompanied with an orthorhombic distortion of the high temperature tetragonal paraelectric $I4/mmm$ parent phase, that lead to polar $Imm2$ phase~\cite{ismailzade78}. Later Needs et $al$.~\cite{needs}, by using the combined refinement of the powder X-ray and neutron diffraction, refined Bi$_2$TiO$_4$F$_2$ and Bi$_2$NbO$_5$F into centrosymmetric tetragonal $I4/mmm$ symmetry and the fluorine anions assigned to the equatorial sites. No evidence of an orthorhombic distortion of the crystal structure was observed. Furthermore, E. McCabe et $al$.~\cite{mccabe2007} showed that Bi$_2$NbO$_5$F adopts a distorted structure with octahedral tilts about [110]$_t$ and [001]$_t$ axis. The absence of SHG signal led them to conclude that the structure was centrosymmetric $Pbca$ and the bond valence sum calculations indicated that fluorine anions prefer the apical sites.
However, all these efforts and numerous investigations have not dispelled, so far, the doubts about ferroelectricity and fluorine ordering in Bi$_2$TiO$_4$F$_2$.

\begin{figure}[t]
\includegraphics[scale=0.60]{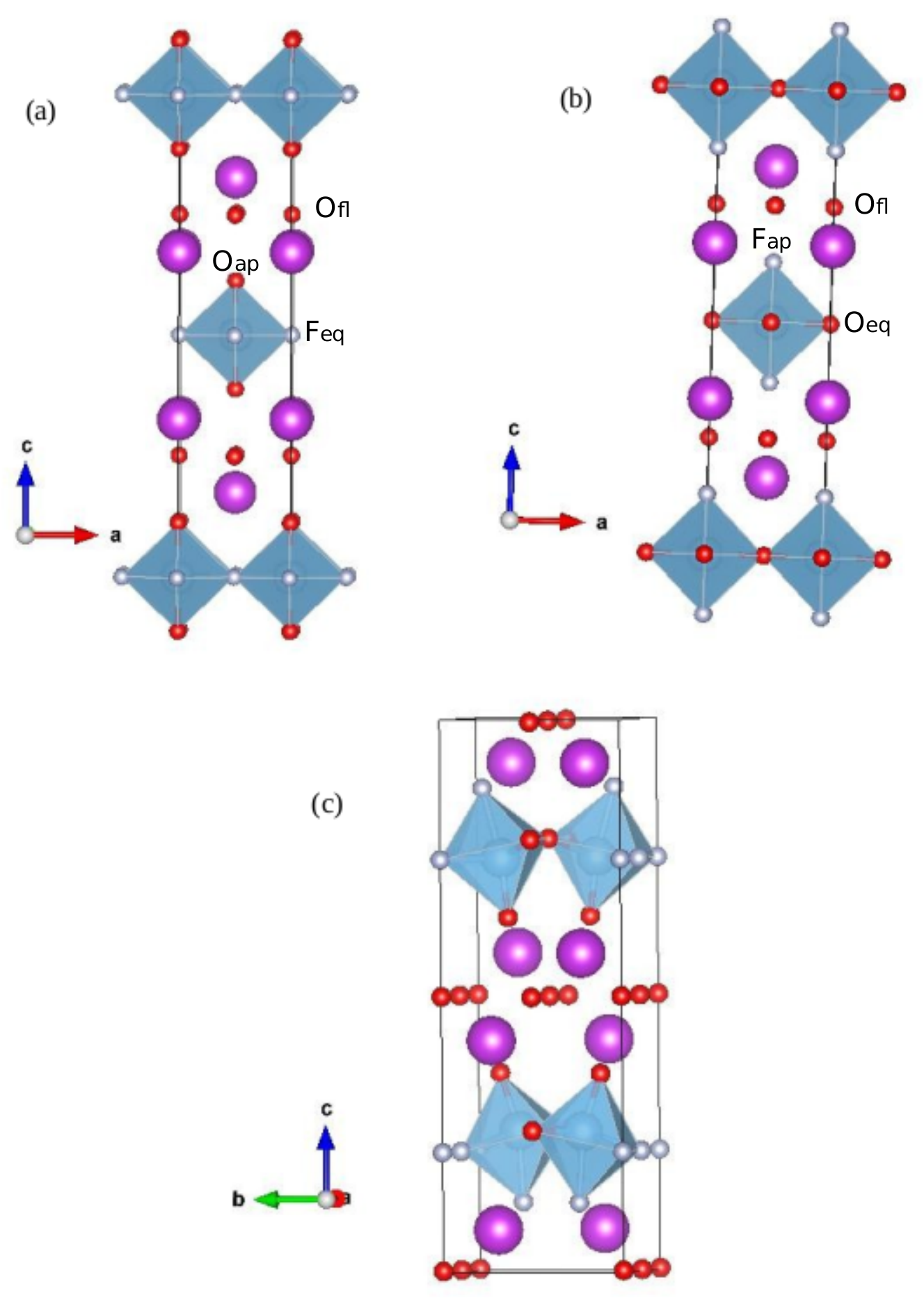}\\
\caption{ High symmetry parent phases structures. (a) Tetragonal $I4/mmm$ (F$_{eq}$), (b) Tetragonal $I4/mmm$ (F$_{ap}$) and (c) Orthorhombic $Pma2$ (F$_{eq}$+F$_{ap}$). Bi atoms in Purple, Ti in blue, F in light grey and O in red } 
\label{figure}
\end{figure}

Here, we have explored different anions ordering configurations for  Bi$_2$TiO$_4$F$_2$ Aurivillius phase, with the aim to examine the effect of fluorine positions on the structural, dielectric and dynamical properties of this compound. We have generated three key configurations: the centrosymmetric tetragonal $I4/mmm$ phase with fluorine anions at the octahedral equatorial sites of the perovskite-like blocks ($I4/mmm$ (F$_{eq}$), see Fig.~\ref{figure}(a)), the centrosymmetric tetragonal $I4/mmm$ phase with fluorine anions at the octahedral apical sites ($I4/mmm$ (F$_{ap}$), see Fig.~\ref{figure}(b)) and the polar orthorhombic  $Pma_2$ (F$_{eq}$+ F$_{ap}$) with $fac$-arrangement, consisting of fluorine anions occupying the three corners of one face of the octahedron (see Fig.~\ref{figure}(c)). Recent calculations have recently found this latter configuration as the most stable among twenty tested ones~\cite{doi:10.1021/acs.jpcc.9b09806}.  

Our study shows that Bi$_2$TiO$_4$F$_2$ adopts a polar ground state of monoclinic $Pc$ symmetry with the fluorine anions distributed over the octahedral equatorial and apical sites (F$_{eq}$+ F$_{ap}$) in a $fac$-arrangement.  Phonon calculations performed in the three different configurations revealed numerous polar, antipolar and octahedral tilts instabilities that lead to many metastable phases and  particularly, a strong polar instability,  along the $c$-axis direction was found in the $I4/mmm$ (F$_{ap}$) configuration.

\section{computational details}
This work have been performed within density functional theory \cite{dft65, dft64} as implemented in ABINIT package \cite{abinit, gonze09, abinit2}. The exchange and correlation functional was evaluated with the generalized gradient approximation GGA-PBEsol as parameterized by perdew, Burke and Ernzerhof~\cite{perdew} where Bi ($5d, 6s, 6p$), Ti ($3d, 4s$), O and F ($2s, 2p$) levels were treated as valence states. Wave functions were expanded up to a kinetic energy cutoff (E$cut$) of 42 Hartrees. Integrals over the Brillouin zone approximated by sums on  4 $\times$ 4 $\times$ 1 Monokhorst-Pack mesh of special $k$ points \cite{monkhors76} was enough for good convergence of structural optimization and phonons calculations (tests on 6 $\times$ 6 $\times$ 1 were also performed for validation). The Broyden-Fletcher-Goldfarb-Shanno \cite{schlegel} minimization algorithm was used to give the structural optimisation.
We relaxed the position of each individual atom until the absolute values of the forces on these atoms were converged to less than $10^{-5}$ Ha/Bohr.
Phonons, Born effective were calculated with a variational approach to density functional perturbation theory (DFPT) \cite{gonze97, baroni01} and the spontaneous polarization was calculated using the Berry phase formalism \cite{resta94}.

\section{Results and Discussion }

As mentioned in the introduction, we generated three high symmetry hypothetical reference structures for Bi$_2$TiO$_4$F$_2$, with different fluorine anions ordering into the octahedral sites: $I4/mmm$ (F$_{eq}$), $I4/mmm$ (F$_{ap}$) and  $Pma_2$ (F$_{eq}$+F$_{ap}$).
cells parameters and atomic coordinates are given in Table \ref{table:1}. After fully relaxation, phonons calculations were performed to identify the kind of instabilities that can lower the internal energy of the different systems.   


\begin{table}[t]
\caption{ Calculated cell parameters for high symmetry tetragonal $I4/mmm$(F$_{eq}$) and $I4/mmm$(F$_{ap}$) and orthorhombic $Pma2$ (F$_{eq}$+F$_{ap}$) phases. Here, the tetragonal phases are doubled along the [110] diagonale $\sqrt2\times\sqrt2\times1$, to allow the atomic positions to match those of the orthorhombic metastable phases. Our results are in good agreement with the experimental cell parameters for the $I4mmm$ phase obtained by Needs and $al$.~\cite{needs} with a=5.37$\AA$ and c=16.32$\AA$ and experimental atomic coordinates between brackets. }
\begin{tabular}{lllllcccccccccccccccrr}
\hline
\hline
Structure&  Atom  &&  x & y & z \\
\hline

\textbf{I4/mmm (F$_\textbf{eq}$})   \\
a=$5.41\mathring{A}$&  Bi & &  0.0000 & 0.0000 & 0.32654\\
c= $16.4\mathring{A}$  && & (0.0000 & 0.0000 & 0.32829)\\
 &   Ti & &  0.0000 & 0.0000 & 0.00000\\ 
\hspace{9pt} 
&  F$_{eq}$ & & 0.2500 & 0.7500 & 0.00000\\
  &O$_{ap}$&& 0.0000 & 0.0000 & 0.11208\\
 &&& (0.0000 & 0.0000 & 0.11614)\\
  &   O$_{fl}$  && 0.2500 & 0.7500 & 0.25000\\
 \hline
\textbf{I4/mmm (F$_\textbf{ap}$})  \\
a=$5.42\mathring{A}$& Bi && 0.0000 & 0.0000 & 0.33317\\
c= $15.2\mathring{A}$  && & (0.0000 & 0.0000 & 0.32829)\\
  &  Ti & & 0.0000 & 0.0000 & 0.00000\\ 
\hspace{9pt}  
 & O$_{eq}$ & & 0.2500 & 0.7500 & 0.00000\\
 & F$_{ap}$ && 0.0000 & 0.0000 & 0.12847\\
 &&  & (0.0000 & 0.0000 & 0.11614)\\
 &  O$_{fl}$ & &  0.7500 & 0.5000 & 0.25000\\
 \hline
\textbf{Pma2 (F$_\textbf{eq}$+F$_\textbf{ap}$)}   \\
a=$5.41\mathring{A}$& Bi$_1$ & & 0.75000 & 0.22394 & 0.41923\\
b=$5.37\mathring{A}$ &    Bi$_2$&&0.75000 & 0.25675 &  0.07642\\
c= $16.25\mathring{A}$ &  Ti && 0.75000 & 0.28640 & 0.74983\\ 
\hspace{9pt}   
&  O$_{eq}$ & & 0.00000 & 0.50000 & 0.25601\\
&    O$_{ap}$   & &   0.25000 & 0.17128 & 0.35827\\
&    O'$_{ap}$   & &    0.25000 & 0.67704 & 0.64230\\
& F$_{eq}$ & &  0.50000 & 0.00000 & 0.21796\\
& F$_{ap}$ & &  0.75000 & 0.18477& 0.87569\\
& F'$_{ap}$ & &  0.75000 & 0.70640& 0.12499\\
&  O$_{fl}$ & & 0.00000 & 0.00000 & 0.49981\\
 \hline
 \hline
\label{table:1}
\end{tabular}
\end{table}

\subsection{High symmetry \textbf{I4/mmm (F$_\textbf{eq}$)} paraelectric phase}

Phonon calculations were performed in this parent phase and numerous mode instabilities were identified at high-symmetry points. They include $(i)$ two [110]$_t$ in-plane polar modes ($\Gamma_{5}^{-}$): one consisting of polar displacements of Bi against O$_{fl}$ and Ti against O$_{ap}$ -Ti and F$_{eq}$ displace in the same direction at variance of what is commonly observed in oxide Aurivillius phases, where B cations displace against O$_{eq}$- ($\Gamma_{5}^{-}$ at 163$i$ cm$^{-1}$) and a second polar rigid layer mode (RL-mode), in which, perovskite blocks displace against (Bi$_2$O$_2$) layers -as commonly observed in  oxide Aurivillius phases- ($\Gamma_{5}^{-}$ at 33$i$ cm$^{-1}$). $(ii)$ [110]$_t$ in-plane antipolar modes (M$_5^-$, M$_5^+$, X$_{3}^{-}$ and X$_{4}^{-}$), consisting of inter-perovskite antipolar motion of O$_{ap}$ (M$_5^-$ at 137$i$ cm$^{-1}$) and inter-perovskite antipolar motion of F$_{eq}$ (M$_5^-$ at 27$i$ cm$^{-1}$); inter-(Bi$_2$O$_2$) layers antipolar motion of Bi (M$_5^+$); antipolar motion of O$_{ap}$ between in-plane linked octahedra (X$_{3}^{-}$ and X$_{4}^{-}$), and $(iii)$ octahedral tilts about [110]$_t$ direction of O$_{ap}$ and F$_{eq}$ (X$_{3}^{+}$ and X$_{4}^{+}$) and about [001] direction of F$_{eq}$ (X$_{2}^{+}$).

Considering the condensation of individual modes, reported in  Table~\ref{table:2}, we see that the octahedral tilt about [110]$_t$ axis, X$_{3}^{+}$ (Cmca) drives the largest energy gain, followed by the polar distortion $\Gamma_{5}^{-}$ ($Fmm2$), the antipolar M$_5^-$ ($Cmcm$) and the octahedral tilt about [100] axis X$_{2}^{+}$, ($Pbam$). M$_5^+$ ($Cmca$) and X$_{4}^{+}$ ($Cccm$) distortions give also a sizeable energy gain. This hierarchy differs from what is noticed in the $m=1$, Bi$_2$W0$_6$ in which the polar $\Gamma_{5}^{-}$ distortion gives the largest energy gain followed by the [001]$_t$ axis tilt,  X$_{2}^{+}$~\cite {djani12}.

Considering the condensation of combined unstable modes, we identified many additional metastable phases at lower energy (see Table~\ref{table:2}). The most stable phase is polar of $Pca2_1$ symmetry, resulting from the combination of  $\Gamma_5^-$, X$_3^+$ and X$_2^+$ ($M_5^+$ mode is brought by trilinear coupling $X_2^+$ $\oplus$ $X_3^+$ $\oplus$ $M_5^+$). This $Pca2_1$ phase is closely followed by polar $Pc$ phase, resulting from the combination of M$_5^-$, X$_3^+$ and $_2^+$ and containing insignificant contributions from M$_5^+$ and $\Gamma_5^-$ mode, not reported in Table~\ref{table:2}. 

The mode-by-mode contributions to the atomic distortion $\Delta$ of each metastable phase with respect to $I4/mmm$ parent reference structure are also reported in table~\ref{table:2}. These contributions are evaluated with qAgate software ~\cite{jordan_bieder_2021_4606005}, by expressing the distortion $\Delta$ in the basis of phonon eigendisplacements vectors $\eta_i$ of the $I4/mmm$ phase (such that $<\eta_i| M | \eta_j> = \delta_{ij}$), following the scheme explained in Ref.~\cite{djani12}: $\Delta = A \sum_i \alpha_i \eta_i$, where $A$ is the total distortion amplitude and $\alpha_i$ are the relative mode contributions such that $\sum_i \alpha_i^2=1$. Contributions of distinct phonon modes $i$ to the distortion $\Delta$ of a given phase correspond therefore to the amplitudes $A \alpha_i$. From  Table~\ref{table:2} (up part), we notice a strong cooperation between $\Gamma_5^{-}$ [163$i$] and $X_3^{+}$ (in $Aba2$) and $M_5^{-}$ [137$i$] and $X_3^{+}$ (in $Pc$), revealed from the distortion amplitudes of these modes, which when combined, became almost the double of those of single ones ( for $\Gamma_5^-$, 7.5 $\mathring{A}$ in $Fmm2$ to 11.8 $\mathring{A}$ in $Aba2$ or 13 $\mathring{A}$ in $Pca2_1$; for $X_3^+$, 11.1 $\mathring{A}$ in $Cmca$ to 22.1 $\mathring{A}$ in $Aba2$; for $M_5^-$, 8.2 $\mathring{A}$ in $Cmcm$ to 12.7 $\mathring{A}$ in $Pc$). These strong modes cooperations lead to the most stable $Pca2_1$ and $Pc$ phases. $Pca2_1$ phase does not exhibit any further phonon instabilities, so can be considered as the hypothetical ground state for $I4/mmm$ (F$_{eq}$) configuration.

\subsection{High symmetry \textbf{I4/mmm (F$_\textbf{ap}$)}  paraelectric phase}

\begin{figure}[]
\includegraphics[scale=0.20]{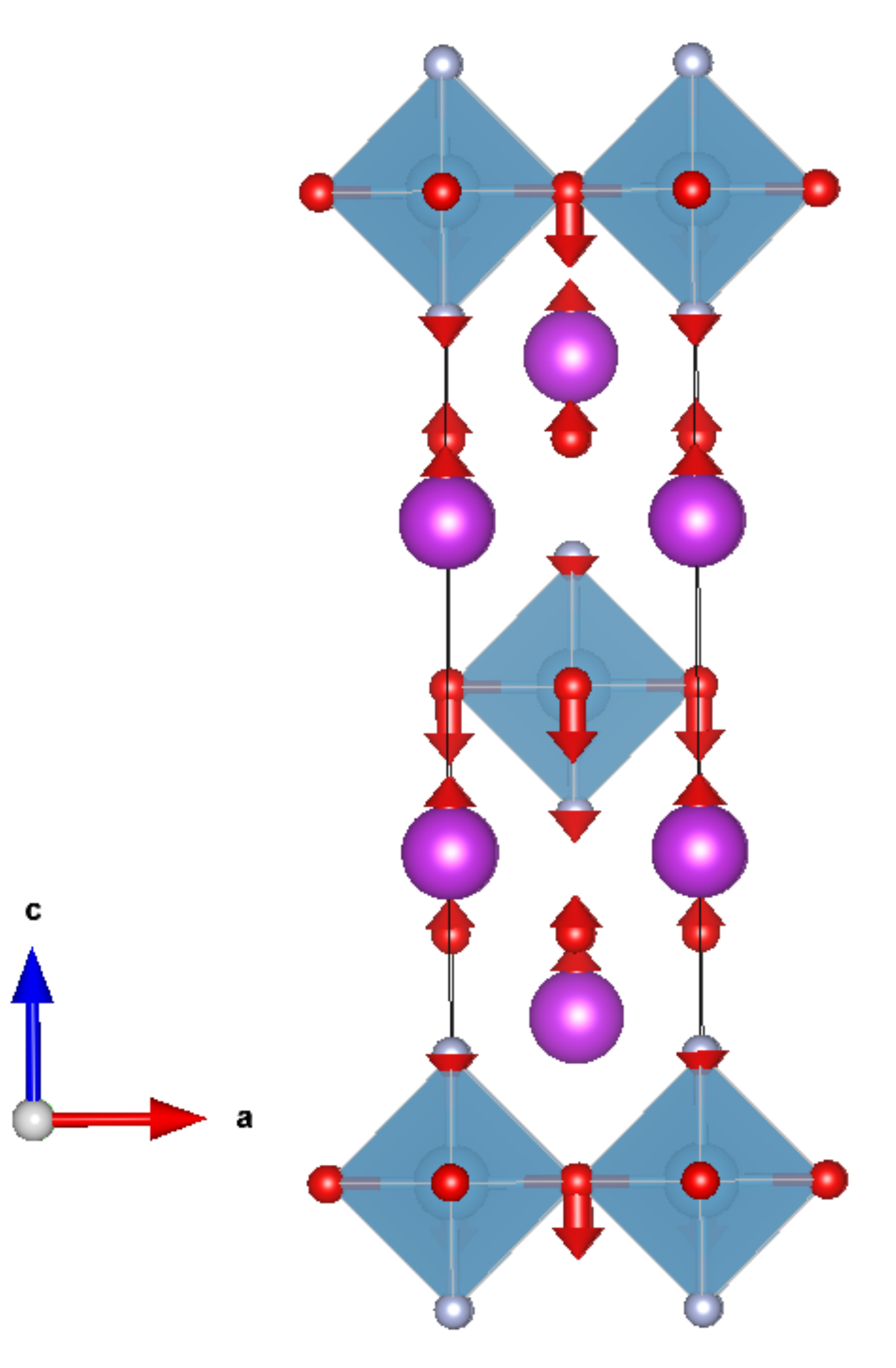}\\
\caption{ Out-of-plane, [001]$_t$ direction polar mode, $\Gamma_3^{-}$. The atomic displacement is showing a Rigid-Layer mode along the polar direction: (Bi$_2$O$_2$) layers displace against Perovskite-like blocks, with a domination of O$_{eq}$ motion against Bi cations. Bi atoms in Purple, Ti in blue, F in light grey and O in red.} 
\label{fig2}
\end{figure}

Phonon calculations performed in the parent $I4/mmm$ (F$_{ap}$) phase with fluorine anions sitting in apical positions, showed the same pattern of unstable modes as frequently found in oxide Aurivillius phases (see~\cite{2bwo}), such as: a polar $\Gamma_5^{-}$ mode (the softest) that consists of polar motion of Ti  against both  O$_{eq}$ and F$_{ap}$ in perovskite blocks and Bi against O$_{fl}$ in (Bi$_2$O$_2$) layers (the Rigid-Layer mode is stable at 66 cm$^{-1}$); an antipolar $M_5^{-}$ mode that consists of inter-perovskite antipolar motion of Ti; out-of-plane octahedral tilt of  O$_{eq}$ and F$_{ap}$ about [110]$_t$ axis (X$_{3}^{+}$ and X$_{4}^{+}$) and in-plane octahedral tilt of O$_{eq}$ about [001]$_t$ axis (X$_{2}^{+}$). In-plane twists of O$_{eq}$ ($\Gamma_4^{-}$ and $M_4^{-}$) were also found. More importantly, an unstable polar mode of $\Gamma_3^{-}$ symmetry, polarized along the out-of-plane [001]$_t$ direction was also found at a frequency of 65$i$ cm$^{-1}$. This mode usually exists in Aurivillius phases, but is never found unstable~\cite{PhysRevB.99.014101} (e.g. at 98 cm$^{-1}$ in $I4/mmm$ (F$_{eq}$), at 88 cm$^{-1}$ in $I4/mmm$ phase of Bi$_2$WO$_6$ and at 65 cm$^{-1}$ in $I4/mmm$ phase of Bi$_2$W$_2$O$_9$). Its eigendisplacement involves mainly a rigid layer motion of the (Bi$_2$O$_2$) layer against the perovskite block along the [001]$_t$ direction, with a domination of Bi motion against O$_{eq}$ (see Fig.~\ref{fig2}). 

  \begin{table*}[ht]
 \caption{Modes contributions $A \alpha_i$  (A in $\mathring{A}$) (see the text) of metastable phase, derived from condensation of individual and combined unstable modes, wrt. paraelectric $I4/mmm$ (F$_{eq}$) (up table) and $I4/mmm$ (F$_{ap}$) (down table) parent phases. Energy difference $\Delta E_{t}$ (in meV/f.u) wrt. paraelectric $I4/mmm$ (F$_{eq}$) (up table) and $I4/mmm$ (F$_{ap}$) is also reported. Frequencies [$\omega$] are in cm$^{-1}$ and Space group are given in a conventional setting.}
\small
\begin{tabularx}{.85\linewidth}{lcccccccccccccccccccccccccccccccccccc}
\hline
\hline
\textbf{I4/mmm (F$_\textbf{eq}$})&&&\multicolumn{2}{c}{\textbf{$\Gamma_5^-$}}&\multicolumn{2}{c}{\textbf{M$_5^-$}}&\textbf{X$_2^+$}&\textbf{X$_3^+$}&\textbf{X$_4^+$}&\textbf{X$_3^-$}&\textbf{M$_5^+$}&\textbf{X$_4^-$}&$\Delta E$\\
phases&irreps&A&[163$i$&33$i$]&[137$i$&27$i$]&[136$i$]&[127$i$]&[118$i$]&[94$i$]&[40$i$]&[30$i$]&\\
\hline
$Fmm2$(42)&\scriptsize $\Gamma_5^-$&11.3&[7.5&8.0]&&&&&&&&&$-$328\\
$Cmcm$(63)&\scriptsize M$_5^-$&10.7&&&[8.2&5.5]&&&&&&&$-$234\\
$Pbam$(55)&\scriptsize X$_2^+$&7.2&&&&&5.2&&&&&&$-$237\\
$Cmca$(64)&\scriptsize X$_3^+$&11.6&&&&&&11.1&&&&&$-$389\\
$Cccm$(66)&\scriptsize X$_4^+$&8.1&&&&&&&7.6&&&&$-$205\\
$Cmcm$(63)&\scriptsize X$_3^-$&10.1&&&&&&&&9.6&&&$-$132\\
$Cmca$(64)&\scriptsize M$_5^+$&29.2&&&&&&&&&29.0&&$-$212\\
$Cmma$(67)&\scriptsize X$_4^-$&1.2&&&&&&&&&&1.2&$-$0.28\\
\\
$Abm2$(39) &\scriptsize$\Gamma_5^{-}$+$X_2^{+}$+ $X_4^{-}$&10.7&[6.4&4.9]&&&4.6&&&&&1.3&$-$423\\
 $Aba2$(41)&\scriptsize $\Gamma_5^{-}$+$X_3^{+}$&29.5&[11.8&5.3]&&&&22.1&&&&&$-$624\\ 
 $P$-$1$ (2)&  \scriptsize $M_5^{+}$+$X_2^{+}$+ $X_3^{+}$&21.8&&&&&3.7&10.3&&&18.5&&$-$467\\
$Pc$ (7)&  \scriptsize $M_5^{-}$+$X_2^{+}$+ $X_3^{+}$&28.3&&&[12.7&6.5]&2.8&15.8&&&&&$-$807\\
 $Pca2_1$(29) & \scriptsize$\Gamma_5^{-}$+ $X_2^{+}$+ $X_3^{+}$+ $M_5^{+}$&32.3&[13.0&5.2]&&&3.5&16.2&&& 15.2&&\textbf{$-$822}\\
 \\
\hline
\end{tabularx}
\begin{tabularx}{.85\linewidth}{lcccccccccccccccccccccccccccccccccccc}
\textbf{I4/mmm (F$_\textbf{ap}$})&&&\multicolumn{1}{c}{\textbf{$\Gamma_5^-$}}&\multicolumn{1}{c}{\textbf{M$_5^-$}}&\multicolumn{1}{c}{\textbf{M$_4^-$}}&\multicolumn{1}{c}{\textbf{$\Gamma_4^-$}}&\multicolumn{1}{c}{\textbf{$X_3^+$}}&\multicolumn{1}{c}{\textbf{X$_2^+$}}&\multicolumn{1}{c}{\textbf{$\Gamma_3^-$}}&\multicolumn{1}{c}{\textbf{X$_4^+$}}&&&&&&&&&$\Delta E$\\

phases&irreps&A&[200$i$]&[197$i$]&[110$i$]&[107$i$]&[99$i$]&[90$i$]&[65$i$]&[54$i$]&&&&&\\

\hline
$Fmm2$(42)&\scriptsize $\Gamma_5^-$&2.2&1.9&&&&&&&&&&&&&&&&$-$27\\
$Cmcm$(63)&\scriptsize M$_5^-$&2.1&&1.8&&&&&&&&&&&&&&&$-$24\\
$P4_2/nmc$(137)&\scriptsize M$_4^-$&2.4&&&2.3&&&&&&&&&&&&&&$-$15\\
$I$-$4m2$(119)&\scriptsize $\Gamma_4^-$&2.3&&&&2.3&&&&&&&&&&&&&$-$14\\
$Cmca$(64)&\scriptsize X$_3^+$&5.4&&&&&5.2&&&&&&&&&&&&$-$56\\
$Pbam$(55)&\scriptsize X$_2^+$&2.8&&&&&&2.0&&&&&&&&&&&$-$14\\
$I4mm$(107)&\scriptsize $\Gamma_3^-$&8.2&&&&&&&8.1&&&&&&&&&&$-$56\\
$Cmcm$(63)&\scriptsize X$_4^+$&3.7&&&&&&&&3.7&&&&&&&&&$-$10\\
 \\
$Abm2$(39)&\scriptsize $\Gamma_5^{-}$+ X$_2^{+}$&3.7&1.9&&&&&2.2&&&&&&&&&&&$-$45\\
$Aba2$(41)&\scriptsize $\Gamma_5^{-}$ + X$_3^{+}$&6.8&2.2&&&&5.9&&&&&&&&&&&&$-$103\\
$Pc$(7)&\scriptsize $\Gamma_5^{-}$+X$_3^+$+X$_2^{+}$+$\Gamma_3^{-}$&9.2&2.2&&&&5.2&0.6&6.4&&&&&&&&&&$-$126\\
\\
$Aba2$(41)&\scriptsize $\Gamma_3^{-}$ + X$_2^{+}$&8.0&&&&&&1.4&7.6&&&&&&&&&&$-$60\\
 $Abm2$(39)&\scriptsize $\Gamma_3^{-}$ + X$_3^{+}$&8.8&&&&&4.7&&7.1&&&&&&&&&&$-$87\\
$Pna2_1$(33)&\scriptsize $\Gamma_3^{-}$+ X$_4^{+}$+M$_5^{-}$&11.4&&2.1&&&&&8.4&5.1&&&&&&&&&$-$114\\
$Pca2_1$(29)&\scriptsize $\Gamma_3^{-}$ + X$_3^{+}$+ X$_2^{+}$+M$_5^{-}$&9.4&&2.2&&&5.3&0.6&6.5&&&&&&&&&&\textbf{$-$124}\\
\hline
\hline
\label{table:2}
\end{tabularx}
\end{table*}

As for $I4/mmm$ (F$_{eq}$), we condensed individual and combined unstable modes into the parent paraelectric phase  $I4/mmm$ (F$_{ap}$) and the results are also reported in Table~\ref{table:2}. It is worth noting that both total amplitude of distortions and energy gain are very small compared to $I4/mmm$ (F$_{eq}$). This is consistent with previous works on Bi$_2$NbO$_5$F compound ~\cite{PhysRevB.48.16061}, that showed that the energy to substitute O$_{ap}$ with fluorine anions is much lower than the energy to substitute O$_{eq}$. The condensation of individual modes gave, equally, the larger energy gain to $\Gamma_3^{-}$ ($I4mm$) and X$_{3}^{+}$ ($Cmca$), followed by polar $\Gamma_{5}^{-}$ ($Fmm2$) and antipolar M$_5^-$ ($Cmcm$). The condensation of combined modes showed that the most stable phase is associated to $\Gamma_3^{-}$ and $\Gamma_5^{-}$ polar modes in addition to X$_{3}^{+}$ and X$_{2}^{+}$ octahedral tilts. This phase of $Pc$ symmetry is polar along the in-plane [110]$_t$ direction and out-of-plane [001]$_t$ direction 
and is considered as the ground state for the (F$_{ap}$) configuration, since further phonon calculations in this phase do not reveal any instabilities. The condensation of polar $\Gamma_3^{-}$, $M_5^{-}$ antipolar mode and X$_{3}^{+}$ and X$_{2}^{+}$ octahedral tilts lead also to a stable phase of $Pca2_1$ symmetry, very close in energy to the ground state $Pc$.

In an aim to explore the origin of out-of-plane $\Gamma_3^{-}$ instability, the ferroelectricity being directy related to the length and covalency of cations-anions bonds~\cite{ravez88}, we examined the selected bonds lengths given in Table~\ref{table:9}. We noticed that F$_{ap}$ is moving away from Ti and getting closer to Bi (Ti-F$_{ap}$=1.95$\mathring{A}$ and Bi-F$_{ap}$=2.77$\mathring{A}$ in $I4/mmm$ (F$_{ap}$), in contrast  to Ti-O$_{ap}$=1.83$\mathring{A}$ and Bi-O$_{ap}$=2.88$\mathring{A}$ in $I4/mmm$ (F$_{eq}$)). Accordingly, the calculations of the Born effective charges ($Z^*$), reported in Table~\ref{table:7}, showed that, 
on the one hand, the ionic character of the Ti-F$_{ap}$ bonding, marked by $Z^{*}_{zz}$ Ti(+4.18$e$) that is close to its nominal value along the apical direction can explain the elongation of the Ti-F$_{ap}$ bonds and the non hybridization between F-$2p$ and Ti-$d$ orbitals and, on the other hand, the strongly anomalous $Z^{*}$ F$_{ap}$(-2.3$e$) and anomalous $Z^{*}$ Bi(+4.2$e$) can be linked to an increasing hybridization between  F-$2p$ and Bi-$6p$-$6s$ orbitals and hence, to a shortening of Bi-F$_{ap}$ bonds.

The ionic character of the Ti-F$_{eq}$ is also confirmed in  $I4/mmm$ (F$_{eq}$) with $Z^{*}_{xx=yy}$ Ti(+4.7$e$), that is close to its nominal value along the equatorial octahedral plane. However, here, the Ti-F$_{eq}$ bond length is equal to Ti-O$_{eq}$ bond length in $I4/mmm$ (F$_{ap}$), although Ti and O$_{eq}$ are highly hybridized ($Z^{*}_{xx=yy}$ Ti(+8.4$e$) and $Z^{*}_{xx=yy}$ O$_{eq}$(-4.3$e$)). This equivalence in bond lengths, despite the clear difference in hybridization between Ti and F$_{eq}$/O$_{eq}$ could be explained by the constrain on bond lengths in the equatorial plane due to the mismatch between the perovskite-like block and the (Bi$_2$O$_2$) layer.


 Moreover, the calculation of the interatomic force constants (IFCs), $C_{\alpha,\beta}(lk,l'k')$, through the expression $F_{\alpha}(lk)=-C_{\alpha,\beta}(lk,l'k')\tau_{\beta}(l'k')$~\cite{PhysRevB.60.836} which relates $\alpha$-component of the force $F_{\alpha}(lk)$ on atom $k$ in cell $l$, to the induced displacement $\tau_{\beta}(l'k')$ of atom $k'$ in cell $l'$, as defined in Refs~\cite{Ghosez_1996}, allow us to identify which driving forces lead the system to structural instabilities~\cite{PhysRevB.60.836,PhysRevB.97.174108}. By convention when the IFC is positive it corresponds to a destabilizing interaction. Accordingly, our results on IFC, reported in Table~\ref{table:10}, showed that the shortening of the Bi-F$_{ap}$ bonds in $I4/mmm$ (F$_{ap}$) configuration decreases the distance between Bi and O$_{eq}$ atoms, thus, increases their hybridization, as reflected by the large and positive IFC of Bi-O$_{eq}$ interaction (+0.011 Ha/Bohr$^2$), in contrast to the IFC of Bi-F$_{eq}$ that is positive but vanishingly small (+0.002 Ha/Bohr).

 \begin{table}[ht]
\begin{center}
\caption{Interatomic bond lengths ($\mathring{A}$) for selected atoms.}
\begin{tabular}{lcccccccccrrrrrrrrrrrrrr}
\hline
\hline
structure&pairs of atoms && distance($\mathring{A}$)\\

\hline
\textbf{I4/mmm (F$_\textbf{eq}$})\\
&Bi-O$_{ap}$& &2.88   \\
&Bi-F$_{eq}$&& 3.42   \\
&Bi-O$_{fl}$&& 2.28\\
&Ti-O$_{ap}$&& 1.83\\
&Ti-F$_{eq}$&&1.91\\
\hline
 \textbf{I4/mmm (F$_\textbf{ap}$})\\
&Bi-F$_{ap}$&& 2.77   \\
&Bi-O$_{eq}$& &3.17   \\
&Bi-O$_{fl}$&& 2.29\\
&Ti-F$_{ap}$&& 1.95\\
&Ti-O$_{eq}$&& 1.91\\
 \hline

\hline
\hline
\end{tabular}
\label{table:9}
\end{center}
\end{table} 

\begin{table}[ht]
\begin{center}
\caption{Nonzero elements of the Born effective charges tensor $Z^*$ (in Cartesian coordinates) for Bi$_2$TiO$_4$F$_2$ in the paraelectric tetragonal $I4/mmm$ phase. The nominal charges of Bi, Ti, O and F are +3, +4, $-$2 and $-$1, respectively.}
\begin{tabular}{llrrrrrrrrrrrrrrrrrrr}
\hline
structure&Atom & $Z_{xx}^{*}$ & $Z_{yy}^{*}$ & $Z_{zz}^{*}$ &$Z_{xy}^{*}$ & $Z_{yx}^{*}$& $Z_{xz}^{*}$ & $Z_{zx}^{*}$\\
\hline
\textbf{I4/mmm (F$_\textbf{eq}$})\\

&Bi & 5.2 & 5.2 & 4.0 &0.0 & 0.0&0.0&0.0\\
&Ti  & 4.7 & 4.7 & 9.0 &0.0 & 0.0&0.0&0.0 \\
&F$_{eq}$ & $-$2.0 & $-$2.0 & $-$0.9 & $-$1.3 & $-$1.3 &0.0&0.0 \\
&O$_{ap}$ & $-$2.4 & $-$2.4 & $-$4.9 &0.0 & 0.0&0.0&0.0 \\
&O$_{fl}$ & $-$3.0 & $-$3.0 & $-$2.6 & 0.0& 0.0&0.0&0.0 \\
\hline
\textbf{I4/mmm (F$_\textbf{ap}$})\\
&Bi & 4.8 & 4.8 & 4.2 &0.0 & 0.0&0.0&0.0 \\
&Ti  & 8.4 & 8.4 & 4.2 &0.0 & 0.0&0.0&0.0 \\
&O$_{eq}$ & $-$4.3 & $-$4.3 & $-$2.0 & $-$2.3& $-$2.3&0.0&0.0 \\
&F$_{ap}$    & $-$1.7 & $-$1.7 & $-$2.3 & 0.0 & 0.0 &0.0&0.0 \\
&O$_{fl}$ & $-$2.9 & $-$2.9 & $-$2.0 & 0.0 & 0.0 &0.0&0.0\\
\hline
\hline
\label{table:7}
\end{tabular}
\end{center}
\end{table}

\begin{table}[ht]
\caption{Interatomic force constants (IFC) in (Ha/$bohr^{2}$) between different selected pairs of atoms in their local coordinates system along the longitudinal ($\parallel$) direction. Dipole-Dipole (DD) and Short-Range (SR) contributions are also reported.}
     \begin{tabular}{lccccccccrrrrrrrrrrrrrr}
\hline
\hline
structure& &Total & DD & SR\\
\hline
\textbf{I4/mmm (F$_\textbf{ap}$})& &&&   \\

& Bi-F$_{ap}$($\parallel$) & 0.011 & 0.013 & $-$0.002 \\
& Bi-O$_{eq}$($\parallel$) & 0.011 & 0.012 & $-$0.001  \\
 
  \hline
\textbf{I4/mmm (F$_\textbf{eq}$}) & &&&   \\

& Bi-O$_{ap}$($\parallel$) &0.015 & 0.022 & $-$0.007  \\
& Bi-F$_{eq}$($\parallel$) & 0.002 & 0.004 & $-$0.002 \\
\hline
\hline
\end{tabular}
\label{table:10}
\end{table}

\subsection{Spontaneous polarization}

In order to estimate the spontaneous polarization P$_s$ for the different metastable polar phases derived from $I4/mmm$ (F$_{eq}$) and $I4/mmm$ (F$_{ap}$) parent phases, we used two methods: from the Berry phase approach~\cite{resta94}, with the careful determination of the polarization quantum (see~\cite{djani12}) and from the knowledge of $Z^{*}$ and atomic displacements following the expression.
\begin{equation*}
    P_{s,\alpha}= \frac{e}{\Omega} \sum_{k, \beta} Z^*_{k, \alpha \beta} \delta \tau_{k, \beta}
\end{equation*}
where $\delta \tau_{k, \beta}$ is the displacement of atom k along direction $\beta$ from the paraelectric to the ferroelectric phase,  $Z^*_{k, \alpha \beta}$the Born effective charge tensor of atom k, and $\Omega$, the unit cell volume.  The results, reported in Table~\ref{table:6} showed
 \begin{table}[ht]
\begin{center}
 \caption{Spontaneous polarization $P_s$ (in $\mu$c/cm$^{2}$) of different metastable polar phases of Bi$_2$TiO$_4$F$_2$. $P_s$(BP) refers to the polarization computed using the Berry-phase approach, and $P_s$($Z^*$) corresponds to the polarization deduced from the knowledge of the Born effective charges and atomic displacements.}
 
\begin{tabular}{lllllllllllllrrrrrrrrrrrr}
\hline
\hline

structure&phase &&direction&  $P_s$($Z^*$) &&  $P_s$(BP)\\
\hline
\textbf{I4/mmm (F$_\textbf{eq}$}) &&&\\
&$Fmm2$(42)  &  &[110]$_t$&    71 & &  72\\
&$Aba2$(41)&&[110]$_t$& 136 & &  138\\
&$Pca2_1$(29) &&[110]$_t$&112  &&110 \\
\hline
\textbf{I4/mmm (F$_\textbf{ap}$}) &&\\

&$Fmm2$(42) &  &[110]$_t$&    34 & &  30\\
&$I4mm$(107) &  &[001]$_t$&  45 & &  41\\
&$Pc$(7)&&[110]$_t$& 39&&44\\
 & &&[001]$_t$& 37&&35\\

\hline
\hline
\label{table:6}
\end{tabular}
\end{center}
\end{table} 
that, in the structure with (F$_{eq}$) configuration, the polar phases display P$_s$ along the [110]$_t$ direction. P$_s$ in $Fmm2$ phase (with only $\Gamma_5^{-}$ polar mode) is smaller than  P$_s$ in $Aba2$ ($\Gamma_5^{-}$ and $X_3^{+}$) and $Pca2_1$ ($\Gamma_5^{-}$, $X_3^{+}$, $X_2^{+}$ and $M_5^{+}$) phases. This  is in contrast to the results obtained in Bi$_2$WO$_6$~\cite{djani12}, in which the polarization decreases from $Fmm2$ to $Aba2$ to $Pca2_1$. The increasing of P$_s$ in $Aba2$ of Bi$_2$TiO$_4$F$_2$ with (F$_{eq}$) is due to the strong cooperation between $\Gamma_5^{-}$ and  $X_3^{+}$ modes, as mentioned in section A, which is weak in the case of Bi$_2$WO$_6$ (see~\cite{djani12}). The polarization can reach, in this case, a very large value of 138 $\mu$c/cm$^{2}$ (see Table~\ref{table:6}).In the structure with (F$_{ap}$) configuration, P$_s$ expands along both [110]$_t$ and [001]$_t$ directions. P$_s$ in $I4mm$ phase (with only $\Gamma_3^{-}$ polar mode) is larger than P$_s$ in $Fmm2$ ($\Gamma_5^{-}$). In this (F$_{ap}$) configuration, the value of P$_s$ is almost constant in the different polar phases, in agreement with the weak cooperative coupling observed between the polar modes and octahedral tilts, as mentioned in section B. Finally, the ground state $Pc$ displays P$_s$ along two directions:44 $\mu$c/cm$^{2}$ along [110]$_t$ direction and 35 $\mu$c/cm$^{2}$ along [001]$_t$ direction. This finding is of valuable importance since ferroelectricity in Aurivillius phases is usually in-plane [110]$_t$ oriented and not very suitable for thin films application, such as nonvolatile memory devices, that would prefer a polarization switching along out-of-plane [001]$_t$ direction to detect the switching states during read and write operations~\cite{auciello}. 




\subsection{High symmetry \textbf{Pma2 (F$_\textbf{eq}$+F$_\textbf{ap}$)}}

 When fluorine anions are distributed over apical and octahedral sites in $fac$-arrangement, the highest symmetry phase is polar with orthorhombic $Pma2$ symmetry,  with always the same sort of anion (O$_{ap}$ or  F$_{ap}$) on both sides of (Bi$_2$O$_2$) layer. 
 Phonon calculations performed in this $Pma2$ phase showed four unstable modes: three modes of $\Gamma_4$ symmetry at frequencies of 68$i$, 61$i$ and 33$i$ cm$^{-1}$ and one mode of $\Gamma_2$ symmetry at 44$i$ cm$^{-1}$. $\Gamma_4$ [68$i$] and $\Gamma_4$ [61$i$] are  octahedral tilts about [001] direction. Because of absence of inversion center and mirror plane perpendicular to [001] direction, octahedral tilts within the two perovskite blocks are decoupled: each $\Gamma_4$ mode defines tilts of octahedra within one single block. $\Gamma_4$ [33$i$] corresponds to a translation of (Bi$_2$O$_2$) layer against O$_{ap}$ and $\Gamma_2$ is an octahedral tilt about [110] direction. 
 
 We reported in Table~\ref{table:5}, the metastable phases arising from the condensation of individual and combined modes. We noticed that the energy gain in this (F$_{ap}$+F$_{eq}$) configuration is very small in comparison to (F$_{ap}$) and (F$_{eq}$) configurations. The condensation of one $\Gamma_4$ mode brought the contribution of the other $\Gamma_4$ modes and the larger energy gain (21 meV/f.u) is attributed to the phase $Pc$ involving the largest contribution of $\Gamma_4$ [61$i$]. Here, we notice, that contrary to $I4/mmm$ (F$_\textbf{eq}$) and $I4/mmm$ (F$_\textbf{ap}$) configurations, the condensation of [001] and [110] octahedral tilts together ($\Gamma_4$ + $\Gamma_2$) in $Pma2$ does not increase the system stability. 
In this case, the ground state is polar of $Pc$ symmetry derived from the condensation of unstable $\Gamma_4$ [001] octahedral tilt mode, but, we were unable to calculate the spontaneous polarization, since no related centrosymmetric parent phase was available in this configuration~\cite{spaldin2012}. 

 
 
  
 
 \begin{table}[ht]
 \caption{Modes contributions $A \alpha_i$  (A in $\mathring{A}$) (see the text) of metastable phase, derived from condensation of individual and combined unstable modes, wrt. $Pma2$ orthorhombic parent phase. Energy difference $\Delta E_{t}$ (in meV/f.u) wrt. $Pma2$ is also reported. Frequencies [$\omega$] are in cm$^{-1}$ and Space group are given in a conventional setting.}
\begin{tabular}{lccccccccccccr}
\hline
\hline
&&&&&\multicolumn{1}{c}{$\Gamma_4$}&\multicolumn{1}{c}{$\Gamma_2$}&&\multicolumn{1}{c}{$\Delta E$}\\
 phase&&&&A&[$i$68 $i$61 $i$33] &[$i$44]&\\ 
\hline
$Pc$(7)&&\scriptsize $\Gamma_4$& &7.6& [2.7 1.9 5.5]&&&$-$17 \\
$Pc$(7)&&\scriptsize $\Gamma_4$& &7.2& [2.7 2.4 5.6]&&&$-$21 \\
$P2$(3)&&\scriptsize $\Gamma_2$&&5.4&&4.9&&$-$11\\
$Pc$(7)&&\scriptsize $\Gamma_4$&& 8.2& [2.3 1.9 6.8]&&&$-$19 \\
\\
$P1$(1)&&\scriptsize $\Gamma_4$+ $\Gamma_2$&&5.8& [-   2.2   -]&4.7&&$-$16 \\
\hline
\hline
\label{table:5}
\end{tabular} 
\end{table}

\begin{figure*}[t]
\includegraphics[scale=0.43]{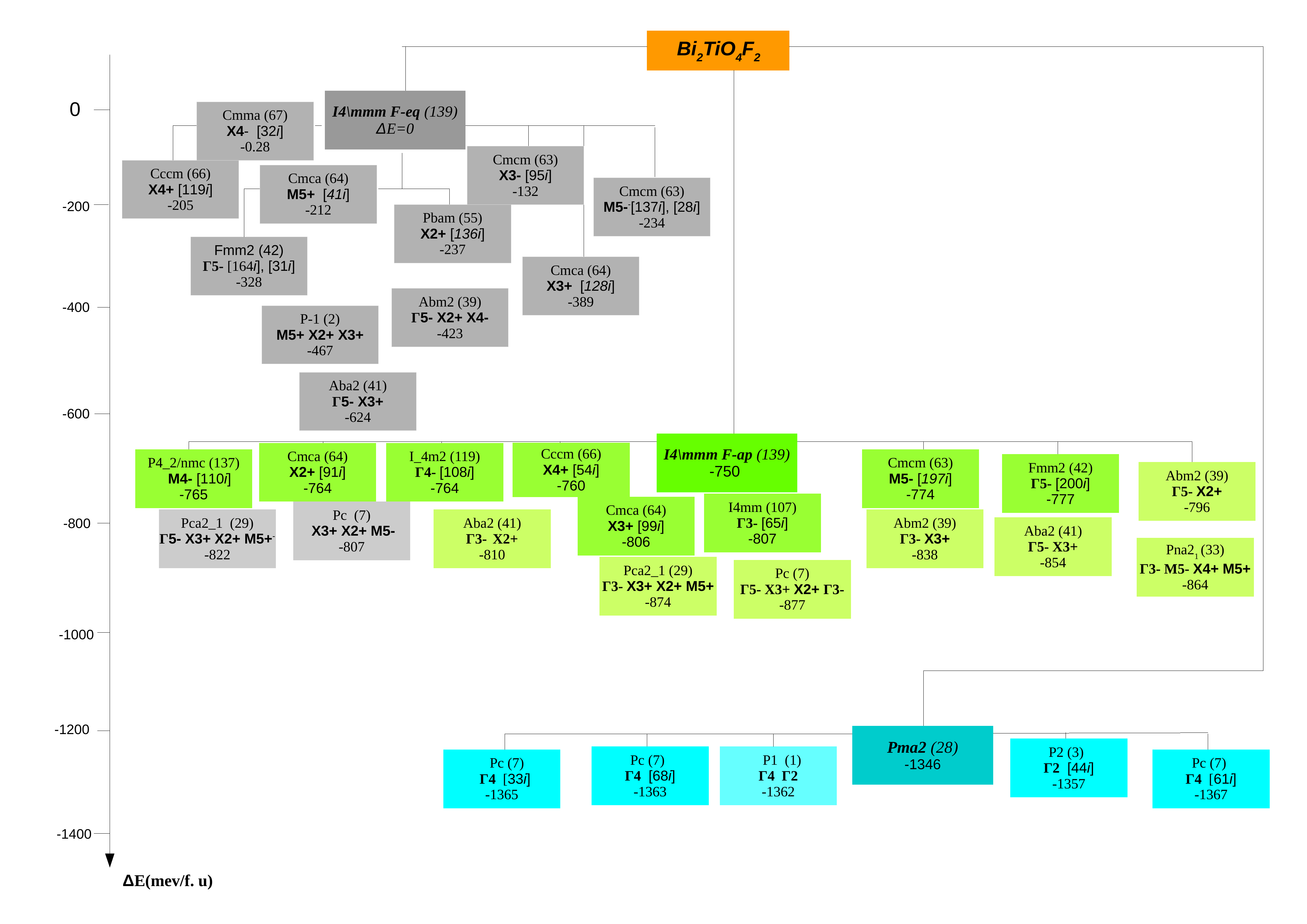} \caption{Summary diagram of relevant metastable phases of Bi$_2$TiO$_4$F$_2$ in different fluorine sites ordering. In the colored boxes are reported the symmetry of the phase (in conventional setting), the involved modes, their irreps and frequencies (between brackets) and $\Delta E_{t}$ (in mev/f.u.) with respect to $I4/mmm$ (F$_{eq}$) taken as the zero energy.}
\label{fig:3}
\end{figure*}

In previous sections, we have discussed independently the key configurations of fluorine sites ordering in Bi$_2$TiO$_4$F$_2$.
In Figure \ref{fig:3}, we provided a global view recapitulation,  comparing the internal energies of the three parent phases and the metastable phases resulting from the condensation of their unstable modes. We found that $I4/mmm$ (F$_{eq}$) is the less stable configuration, followed by $I4/mmm$ (F$_{ap}$) to finally identify the orthorhombic $Pma2$ (F$_{ap}$+F$_{eq}$) as the most stable configuration (consistently with the calculations of K. Morita ~\cite{doi:10.1021/acs.jpcc.9b09806}) and its derived polar monoclinic $Pc$ phase as the most stable among the different metastables phases explored herein.
Our calculations do not fit the $I4/mmm$ experimental model of Needs et $al$. ~\cite{needs} with fluorine assigned to equatorial sites thanks to Bond Valence Sum calculations. However, it is important to point out that these authors mentioned the possibility of symmetry lowering to polar $I4mm$, via a phase transition occuring with a very small perturbation that is not apparent to diffraction techniques~\cite{B401787K, B006120O, B102092G, HYATT2003837}. In particular, these authors exclude the possible assignment of an equal mixture of oxide and fluorine anions over apical and equatorial sites, although their BVS calculations showed close values between those on mixture of fluorine and oxide anions and those on fluorine at equatorial sites only. This excluded assignment is consistent with our (F$_{ap}$+F$_{eq}$) $fac$-arrangement.

\section{conclusion}
We have shown in the $m$=1 oxyfluoride Aurivillius phase Bi$_2$TiO$_4$F$_2$, that the ground state is polar of $Pc$ symmetry and that the ordered distribution of fluorine anions over $both$ apical and equatorial octahedral sites in the perovskite-like blocks (F$_\textbf{eq}$+F$_\textbf{ap}$) is favored. When fluorine anions are positioned at equatorial sites only (F$_\textbf{eq}$), the most favorable phase is polar of $Pca2_1$ symmetry, deriving from small distortions of the parent $I4/mmm$ phase, involving in-plane polar and antipolar displacements of Bi and Ti cations and in-plane and out-of-plane octahedral tilts. Whereas, when fluorine anions are at apical sites only  (F$_\textbf{ap}$), the most favorable phase is polar of $Pc$ symmetry, also deriving from parent $I4/mmm$ phase and involving in-plane polar displacement of Bi and Ti cations, out-of-plane polar displacement of Bi against O$_{eq}$ and in-plane and out-of-plane octahedral tilts.
In particular, this $Pc$ phase is displaying two distinct spontaneous polarizations: in-plane P$_s$= 44 $\mu$c/cm$^{2}$ and out-of-plane P$_s$= 35 $\mu$c/cm$^{2}$. This phase is of great importance in FeRAM's thin films applications and appropriate strain engineering on this phase should be investigated to increase its stability.

\bibliography{library}

\begin{thebibliography}{39}
\expandafter\ifx\csname natexlab\endcsname\relax\def\natexlab#1{#1}\fi
\expandafter\ifx\csname bibnamefont\endcsname\relax
  \def\bibnamefont#1{#1}\fi
\expandafter\ifx\csname bibfnamefont\endcsname\relax
  \def\bibfnamefont#1{#1}\fi
\expandafter\ifx\csname citenamefont\endcsname\relax
  \def\citenamefont#1{#1}\fi
\expandafter\ifx\csname url\endcsname\relax
  \def\url#1{\texttt{#1}}\fi
\expandafter\ifx\csname urlprefix\endcsname\relax\def\urlprefix{URL }\fi
\providecommand{\bibinfo}[2]{#2}
\providecommand{\eprint}[2][]{\url{#2}}

\bibitem[{\citenamefont{Aurivillius}(1949)}]{aurivillius49}
\bibinfo{author}{\bibfnamefont{B.}~\bibnamefont{Aurivillius}},
  \bibinfo{journal}{Arki for Kemi} \textbf{\bibinfo{volume}{1}},
  \bibinfo{pages}{463} (\bibinfo{year}{1949}).

\bibitem[{\citenamefont{Perez-Mato et~al.}(2008)\citenamefont{Perez-Mato,
  Blaha, Schwarz, Aroyo, Orobengoa, Etxebarria, and
  Garc\'{\i}a}}]{perez-mato08}
\bibinfo{author}{\bibfnamefont{J.~M.} \bibnamefont{Perez-Mato}},
  \bibinfo{author}{\bibfnamefont{P.}~\bibnamefont{Blaha}},
  \bibinfo{author}{\bibfnamefont{K.}~\bibnamefont{Schwarz}},
  \bibinfo{author}{\bibfnamefont{M.}~\bibnamefont{Aroyo}},
  \bibinfo{author}{\bibfnamefont{D.}~\bibnamefont{Orobengoa}},
  \bibinfo{author}{\bibfnamefont{I.}~\bibnamefont{Etxebarria}},
  \bibnamefont{and}
  \bibinfo{author}{\bibfnamefont{A.}~\bibnamefont{Garc\'{\i}a}},
  \bibinfo{journal}{Phys. Rev. B} \textbf{\bibinfo{volume}{77}},
  \bibinfo{pages}{184104} (\bibinfo{year}{2008}),
  \urlprefix\url{https://link.aps.org/doi/10.1103/PhysRevB.77.184104}.

\bibitem[{\citenamefont{Withers et~al.}(1991)\citenamefont{Withers, Thompson,
  and Rae}}]{withers91}
\bibinfo{author}{\bibfnamefont{R.~L.} \bibnamefont{Withers}},
  \bibinfo{author}{\bibfnamefont{J.~G.} \bibnamefont{Thompson}},
  \bibnamefont{and} \bibinfo{author}{\bibfnamefont{A.~D.} \bibnamefont{Rae}},
  \bibinfo{journal}{J.\ Solid State Chem.} \textbf{\bibinfo{volume}{94}},
  \bibinfo{pages}{404} (\bibinfo{year}{1991}).

\bibitem[{\citenamefont{Perez-Mato et~al.}(2004)\citenamefont{Perez-Mato,
  Aroyo, Garc\'{i}a, Blaha, Schwarz, Schweifer, and Parlinski}}]{perez-mato04}
\bibinfo{author}{\bibfnamefont{J.~M.} \bibnamefont{Perez-Mato}},
  \bibinfo{author}{\bibfnamefont{M.}~\bibnamefont{Aroyo}},
  \bibinfo{author}{\bibfnamefont{A.}~\bibnamefont{Garc\'{i}a}},
  \bibinfo{author}{\bibfnamefont{P.}~\bibnamefont{Blaha}},
  \bibinfo{author}{\bibfnamefont{K.}~\bibnamefont{Schwarz}},
  \bibinfo{author}{\bibfnamefont{J.}~\bibnamefont{Schweifer}},
  \bibnamefont{and}
  \bibinfo{author}{\bibfnamefont{K.}~\bibnamefont{Parlinski}},
  \bibinfo{journal}{Phys.\ Rev.\ B} \textbf{\bibinfo{volume}{70}},
  \bibinfo{pages}{214111} (\bibinfo{year}{2004}).

\bibitem[{\citenamefont{Etxebarria et~al.}(2010)\citenamefont{Etxebarria,
  Perez-Mato, and Boullay}}]{etxebarria}
\bibinfo{author}{\bibfnamefont{I.}~\bibnamefont{Etxebarria}},
  \bibinfo{author}{\bibfnamefont{J.~M.} \bibnamefont{Perez-Mato}},
  \bibnamefont{and} \bibinfo{author}{\bibfnamefont{P.}~\bibnamefont{Boullay}},
  \bibinfo{journal}{Ferroelectrics} \textbf{\bibinfo{volume}{401}},
  \bibinfo{pages}{17} (\bibinfo{year}{2010}).

\bibitem[{\citenamefont{Hervoches et~al.}(2002)\citenamefont{Hervoches,
  Snedden, Riggs, Kilcoyne, Manuel, and Lightfoot}}]{hervoches}
\bibinfo{author}{\bibfnamefont{C.~H.} \bibnamefont{Hervoches}},
  \bibinfo{author}{\bibfnamefont{A.}~\bibnamefont{Snedden}},
  \bibinfo{author}{\bibfnamefont{R.}~\bibnamefont{Riggs}},
  \bibinfo{author}{\bibfnamefont{S.~H.} \bibnamefont{Kilcoyne}},
  \bibinfo{author}{\bibfnamefont{P.}~\bibnamefont{Manuel}}, \bibnamefont{and}
  \bibinfo{author}{\bibfnamefont{P.}~\bibnamefont{Lightfoot}},
  \bibinfo{journal}{Journal of Solid State Chemistry}
  \textbf{\bibinfo{volume}{164}}, \bibinfo{pages}{280 } (\bibinfo{year}{2002}),
  ISSN \bibinfo{issn}{0022-4596},
  \urlprefix\url{http://www.sciencedirect.com/science/article/pii/S0022459601994733}.

\bibitem[{\citenamefont{Aurivillius}(1953)}]{aurivillius53}
\bibinfo{author}{\bibfnamefont{B.}~\bibnamefont{Aurivillius}},
  \bibinfo{journal}{Arkiv for kemi} \textbf{\bibinfo{volume}{5}},
  \bibinfo{pages}{39} (\bibinfo{year}{1953}).

\bibitem[{\citenamefont{Wang et~al.}(2015)\citenamefont{Wang, Mao, Ye, and
  Huang}}]{C5RA14288A}
\bibinfo{author}{\bibfnamefont{W.}~\bibnamefont{Wang}},
  \bibinfo{author}{\bibfnamefont{W.}~\bibnamefont{Mao}},
  \bibinfo{author}{\bibfnamefont{Z.}~\bibnamefont{Ye}}, \bibnamefont{and}
  \bibinfo{author}{\bibfnamefont{J.}~\bibnamefont{Huang}},
  \bibinfo{journal}{RSC Adv.} \textbf{\bibinfo{volume}{5}},
  \bibinfo{pages}{81087} (\bibinfo{year}{2015}),
  \urlprefix\url{http://dx.doi.org/10.1039/C5RA14288A}.

\bibitem[{\citenamefont{Wang et~al.}(2011)\citenamefont{Wang, Huang, Wang, Liu,
  Wei, Qin, Zhang, and Dai}}]{C1DT10889A}
\bibinfo{author}{\bibfnamefont{S.}~\bibnamefont{Wang}},
  \bibinfo{author}{\bibfnamefont{B.}~\bibnamefont{Huang}},
  \bibinfo{author}{\bibfnamefont{Z.}~\bibnamefont{Wang}},
  \bibinfo{author}{\bibfnamefont{Y.}~\bibnamefont{Liu}},
  \bibinfo{author}{\bibfnamefont{W.}~\bibnamefont{Wei}},
  \bibinfo{author}{\bibfnamefont{X.}~\bibnamefont{Qin}},
  \bibinfo{author}{\bibfnamefont{X.}~\bibnamefont{Zhang}}, \bibnamefont{and}
  \bibinfo{author}{\bibfnamefont{Y.}~\bibnamefont{Dai}},
  \bibinfo{journal}{Dalton Trans.} \textbf{\bibinfo{volume}{40}},
  \bibinfo{pages}{12670} (\bibinfo{year}{2011}),
  \urlprefix\url{http://dx.doi.org/10.1039/C1DT10889A}.

\bibitem[{\citenamefont{Ismailzade and Ravez}(1978)}]{ismailzade78}
\bibinfo{author}{\bibfnamefont{I.~H.} \bibnamefont{Ismailzade}}
  \bibnamefont{and} \bibinfo{author}{\bibfnamefont{J.}~\bibnamefont{Ravez}},
  \bibinfo{journal}{Ferroelectrics} \textbf{\bibinfo{volume}{21}},
  \bibinfo{pages}{423} (\bibinfo{year}{1978}),
  \eprint{https://doi.org/10.1080/00150197808237285},
  \urlprefix\url{https://doi.org/10.1080/00150197808237285}.

\bibitem[{\citenamefont{Needs et~al.}(2005)\citenamefont{Needs, Dann, Weller,
  Cherryman, and Harris}}]{needs}
\bibinfo{author}{\bibfnamefont{R.~L.} \bibnamefont{Needs}},
  \bibinfo{author}{\bibfnamefont{S.~E.} \bibnamefont{Dann}},
  \bibinfo{author}{\bibfnamefont{M.~T.} \bibnamefont{Weller}},
  \bibinfo{author}{\bibfnamefont{J.~C.} \bibnamefont{Cherryman}},
  \bibnamefont{and} \bibinfo{author}{\bibfnamefont{R.~K.}
  \bibnamefont{Harris}}, \bibinfo{journal}{J. Mater. Chem.}
  \textbf{\bibinfo{volume}{15}}, \bibinfo{pages}{2399} (\bibinfo{year}{2005}),
  \urlprefix\url{http://dx.doi.org/10.1039/B502499D}.

\bibitem[{\citenamefont{McCabe et~al.}(2007)\citenamefont{McCabe, Jones, Zhang,
  Hyatt, and Greaves}}]{mccabe2007}
\bibinfo{author}{\bibfnamefont{E.~E.} \bibnamefont{McCabe}},
  \bibinfo{author}{\bibfnamefont{I.~P.} \bibnamefont{Jones}},
  \bibinfo{author}{\bibfnamefont{D.}~\bibnamefont{Zhang}},
  \bibinfo{author}{\bibfnamefont{N.~C.} \bibnamefont{Hyatt}}, \bibnamefont{and}
  \bibinfo{author}{\bibfnamefont{C.}~\bibnamefont{Greaves}},
  \bibinfo{journal}{J. Mater. Chem.} \textbf{\bibinfo{volume}{17}},
  \bibinfo{pages}{1193} (\bibinfo{year}{2007}),
  \urlprefix\url{http://dx.doi.org/10.1039/B613970A}.

\bibitem[{\citenamefont{Morita et~al.}(2019)\citenamefont{Morita, Park, Kim,
  Yasuoka, and Walsh}}]{doi:10.1021/acs.jpcc.9b09806}
\bibinfo{author}{\bibfnamefont{K.}~\bibnamefont{Morita}},
  \bibinfo{author}{\bibfnamefont{J.-S.} \bibnamefont{Park}},
  \bibinfo{author}{\bibfnamefont{S.}~\bibnamefont{Kim}},
  \bibinfo{author}{\bibfnamefont{K.}~\bibnamefont{Yasuoka}}, \bibnamefont{and}
  \bibinfo{author}{\bibfnamefont{A.}~\bibnamefont{Walsh}},
  \bibinfo{journal}{The Journal of Physical Chemistry C}
  \textbf{\bibinfo{volume}{123}}, \bibinfo{pages}{29155}
  (\bibinfo{year}{2019}), \eprint{https://doi.org/10.1021/acs.jpcc.9b09806},
  \urlprefix\url{https://doi.org/10.1021/acs.jpcc.9b09806}.

\bibitem[{\citenamefont{Kohn and Sham}(1965)}]{dft65}
\bibinfo{author}{\bibfnamefont{W.}~\bibnamefont{Kohn}} \bibnamefont{and}
  \bibinfo{author}{\bibfnamefont{L.~J.} \bibnamefont{Sham}},
  \bibinfo{journal}{Phys. Rev.} \textbf{\bibinfo{volume}{140}},
  \bibinfo{pages}{A1133} (\bibinfo{year}{1965}),
  \urlprefix\url{https://link.aps.org/doi/10.1103/PhysRev.140.A1133}.

\bibitem[{\citenamefont{Hohenberg and Kohn}(1964)}]{dft64}
\bibinfo{author}{\bibfnamefont{P.}~\bibnamefont{Hohenberg}} \bibnamefont{and}
  \bibinfo{author}{\bibfnamefont{W.}~\bibnamefont{Kohn}},
  \bibinfo{journal}{Phys. Rev.} \textbf{\bibinfo{volume}{136}},
  \bibinfo{pages}{B864} (\bibinfo{year}{1964}),
  \urlprefix\url{https://link.aps.org/doi/10.1103/PhysRev.136.B864}.

\bibitem[{\citenamefont{Gonze et~al.}(2002)\citenamefont{Gonze, Beuken,
  Caracas, Detraux, Fuchs, Rignanese, Sindic, Verstraete, Zerah, Jollet
  et~al.}}]{abinit}
\bibinfo{author}{\bibfnamefont{X.}~\bibnamefont{Gonze}},
  \bibinfo{author}{\bibfnamefont{J.-M.} \bibnamefont{Beuken}},
  \bibinfo{author}{\bibfnamefont{R.}~\bibnamefont{Caracas}},
  \bibinfo{author}{\bibfnamefont{F.}~\bibnamefont{Detraux}},
  \bibinfo{author}{\bibfnamefont{M.}~\bibnamefont{Fuchs}},
  \bibinfo{author}{\bibfnamefont{G.-M.} \bibnamefont{Rignanese}},
  \bibinfo{author}{\bibfnamefont{L.}~\bibnamefont{Sindic}},
  \bibinfo{author}{\bibfnamefont{M.}~\bibnamefont{Verstraete}},
  \bibinfo{author}{\bibfnamefont{G.}~\bibnamefont{Zerah}},
  \bibinfo{author}{\bibfnamefont{F.}~\bibnamefont{Jollet}},
  \bibnamefont{et~al.}, \bibinfo{journal}{Computational Materials Science}
  \textbf{\bibinfo{volume}{25}}, \bibinfo{pages}{478 } (\bibinfo{year}{2002}),
  ISSN \bibinfo{issn}{0927-0256},
  \urlprefix\url{http://www.sciencedirect.com/science/article/pii/S0927025602003257}.

\bibitem[{\citenamefont{Gonze et~al.}(2009)\citenamefont{Gonze, Amadon,
  Anglade, Beuken, Bottin, Boulanger, Bruneval, Caliste, Caracas, Côté
  et~al.}}]{gonze09}
\bibinfo{author}{\bibfnamefont{X.}~\bibnamefont{Gonze}},
  \bibinfo{author}{\bibfnamefont{B.}~\bibnamefont{Amadon}},
  \bibinfo{author}{\bibfnamefont{P.-M.} \bibnamefont{Anglade}},
  \bibinfo{author}{\bibfnamefont{J.-M.} \bibnamefont{Beuken}},
  \bibinfo{author}{\bibfnamefont{F.}~\bibnamefont{Bottin}},
  \bibinfo{author}{\bibfnamefont{P.}~\bibnamefont{Boulanger}},
  \bibinfo{author}{\bibfnamefont{F.}~\bibnamefont{Bruneval}},
  \bibinfo{author}{\bibfnamefont{D.}~\bibnamefont{Caliste}},
  \bibinfo{author}{\bibfnamefont{R.}~\bibnamefont{Caracas}},
  \bibinfo{author}{\bibfnamefont{M.}~\bibnamefont{Côté}},
  \bibnamefont{et~al.}, \bibinfo{journal}{Computer Physics Communications}
  \textbf{\bibinfo{volume}{180}}, \bibinfo{pages}{2582 }
  (\bibinfo{year}{2009}), ISSN \bibinfo{issn}{0010-4655}, \bibinfo{note}{40
  YEARS OF CPC: A celebratory issue focused on quality software for high
  performance, grid and novel computing architectures},
  \urlprefix\url{http://www.sciencedirect.com/science/article/pii/S0010465509002276}.

\bibitem[{\citenamefont{Gonze et~al.}(2005)\citenamefont{Gonze, Rignanese,
  Verstraete, Beuken, Pouillon, Caracas, Jollet, Torrent, Zerah, Mikami
  et~al.}}]{abinit2}
\bibinfo{author}{\bibfnamefont{X.}~\bibnamefont{Gonze}},
  \bibinfo{author}{\bibfnamefont{G.}~\bibnamefont{Rignanese}},
  \bibinfo{author}{\bibfnamefont{M.}~\bibnamefont{Verstraete}},
  \bibinfo{author}{\bibfnamefont{J.}~\bibnamefont{Beuken}},
  \bibinfo{author}{\bibfnamefont{Y.}~\bibnamefont{Pouillon}},
  \bibinfo{author}{\bibfnamefont{R.}~\bibnamefont{Caracas}},
  \bibinfo{author}{\bibfnamefont{F.}~\bibnamefont{Jollet}},
  \bibinfo{author}{\bibfnamefont{M.}~\bibnamefont{Torrent}},
  \bibinfo{author}{\bibfnamefont{G.}~\bibnamefont{Zerah}},
  \bibinfo{author}{\bibfnamefont{M.}~\bibnamefont{Mikami}},
  \bibnamefont{et~al.}, \bibinfo{journal}{Zeitschrift für Kristallographie}
  \textbf{\bibinfo{volume}{220}}, \bibinfo{pages}{558} (\bibinfo{year}{2005}),
  ISSN \bibinfo{issn}{0044-2968}, \bibinfo{note}{© 2005 Oldenbourg
  Wissenschaftsverlag. Published in Zeitschrift für Kristallographie and
  uploaded in accordance with the publisher's self archiving policy.}

\bibitem[{\citenamefont{Perdew et~al.}(1996)\citenamefont{Perdew, Burke, and
  Ernzerhof}}]{perdew}
\bibinfo{author}{\bibfnamefont{J.~P.} \bibnamefont{Perdew}},
  \bibinfo{author}{\bibfnamefont{K.}~\bibnamefont{Burke}}, \bibnamefont{and}
  \bibinfo{author}{\bibfnamefont{M.}~\bibnamefont{Ernzerhof}},
  \bibinfo{journal}{Phys. Rev. Lett.} \textbf{\bibinfo{volume}{77}},
  \bibinfo{pages}{3865} (\bibinfo{year}{1996}),
  \urlprefix\url{https://link.aps.org/doi/10.1103/PhysRevLett.77.3865}.

\bibitem[{\citenamefont{Monkhorst and Pack}(1976)}]{monkhors76}
\bibinfo{author}{\bibfnamefont{H.~J.} \bibnamefont{Monkhorst}}
  \bibnamefont{and} \bibinfo{author}{\bibfnamefont{J.~D.} \bibnamefont{Pack}},
  \bibinfo{journal}{Phys. Rev. B} \textbf{\bibinfo{volume}{13}},
  \bibinfo{pages}{5188} (\bibinfo{year}{1976}),
  \urlprefix\url{https://link.aps.org/doi/10.1103/PhysRevB.13.5188}.

\bibitem[{\citenamefont{Schlegel}(1982)}]{schlegel}
\bibinfo{author}{\bibfnamefont{H.~B.} \bibnamefont{Schlegel}},
  \bibinfo{journal}{The Journal of Chemical Physics}
  \textbf{\bibinfo{volume}{77}}, \bibinfo{pages}{3676} (\bibinfo{year}{1982}),
  \eprint{https://doi.org/10.1063/1.444270},
  \urlprefix\url{https://doi.org/10.1063/1.444270}.

\bibitem[{\citenamefont{Gonze and Lee}(1997)}]{gonze97}
\bibinfo{author}{\bibfnamefont{X.}~\bibnamefont{Gonze}} \bibnamefont{and}
  \bibinfo{author}{\bibfnamefont{C.}~\bibnamefont{Lee}},
  \bibinfo{journal}{Phys. Rev. B} \textbf{\bibinfo{volume}{55}},
  \bibinfo{pages}{10355} (\bibinfo{year}{1997}),
  \urlprefix\url{https://link.aps.org/doi/10.1103/PhysRevB.55.10355}.

\bibitem[{\citenamefont{Baroni et~al.}(2001)\citenamefont{Baroni, de~Gironcoli,
  Dal~Corso, and Giannozzi}}]{baroni01}
\bibinfo{author}{\bibfnamefont{S.}~\bibnamefont{Baroni}},
  \bibinfo{author}{\bibfnamefont{S.}~\bibnamefont{de~Gironcoli}},
  \bibinfo{author}{\bibfnamefont{A.}~\bibnamefont{Dal~Corso}},
  \bibnamefont{and}
  \bibinfo{author}{\bibfnamefont{P.}~\bibnamefont{Giannozzi}},
  \bibinfo{journal}{Rev. Mod. Phys.} \textbf{\bibinfo{volume}{73}},
  \bibinfo{pages}{515} (\bibinfo{year}{2001}),
  \urlprefix\url{https://link.aps.org/doi/10.1103/RevModPhys.73.515}.

\bibitem[{\citenamefont{Resta}(1994)}]{resta94}
\bibinfo{author}{\bibfnamefont{R.}~\bibnamefont{Resta}}, \bibinfo{journal}{Rev.
  Mod. Phys.} \textbf{\bibinfo{volume}{66}}, \bibinfo{pages}{899}
  (\bibinfo{year}{1994}),
  \urlprefix\url{https://link.aps.org/doi/10.1103/RevModPhys.66.899}.

\bibitem[{\citenamefont{Djani et~al.}(2012)\citenamefont{Djani, Bousquet,
  Kellou, and Ghosez}}]{djani12}
\bibinfo{author}{\bibfnamefont{H.}~\bibnamefont{Djani}},
  \bibinfo{author}{\bibfnamefont{E.}~\bibnamefont{Bousquet}},
  \bibinfo{author}{\bibfnamefont{A.}~\bibnamefont{Kellou}}, \bibnamefont{and}
  \bibinfo{author}{\bibfnamefont{P.}~\bibnamefont{Ghosez}},
  \bibinfo{journal}{Phys. Rev. B} \textbf{\bibinfo{volume}{86}},
  \bibinfo{pages}{054107} (\bibinfo{year}{2012}),
  \urlprefix\url{http://link.aps.org/doi/10.1103/PhysRevB.86.054107}.

\bibitem[{\citenamefont{Bieder}(2021)}]{jordan_bieder_2021_4606005}
\bibinfo{author}{\bibfnamefont{J.}~\bibnamefont{Bieder}},
  \emph{\bibinfo{title}{Qt interface for agate}} (\bibinfo{year}{2021}),
  \urlprefix\url{https://doi.org/10.5281/zenodo.4606005}.

\bibitem[{\citenamefont{Djani et~al.}(2020)\citenamefont{Djani, McCabe, Zhang,
  Halasyamani, Feteira, Bieder, Bousquet, and Ghosez}}]{2bwo}
\bibinfo{author}{\bibfnamefont{H.}~\bibnamefont{Djani}},
  \bibinfo{author}{\bibfnamefont{E.~E.} \bibnamefont{McCabe}},
  \bibinfo{author}{\bibfnamefont{W.}~\bibnamefont{Zhang}},
  \bibinfo{author}{\bibfnamefont{P.~S.} \bibnamefont{Halasyamani}},
  \bibinfo{author}{\bibfnamefont{A.}~\bibnamefont{Feteira}},
  \bibinfo{author}{\bibfnamefont{J.}~\bibnamefont{Bieder}},
  \bibinfo{author}{\bibfnamefont{E.}~\bibnamefont{Bousquet}}, \bibnamefont{and}
  \bibinfo{author}{\bibfnamefont{P.}~\bibnamefont{Ghosez}},
  \bibinfo{journal}{Phys. Rev. B} \textbf{\bibinfo{volume}{101}},
  \bibinfo{pages}{134113} (\bibinfo{year}{2020}),
  \urlprefix\url{https://link.aps.org/doi/10.1103/PhysRevB.101.134113}.

\bibitem[{\citenamefont{Co et~al.}(2019)\citenamefont{Co, Sun, Alpay, and
  Nayak}}]{PhysRevB.99.014101}
\bibinfo{author}{\bibfnamefont{K.}~\bibnamefont{Co}},
  \bibinfo{author}{\bibfnamefont{F.-C.} \bibnamefont{Sun}},
  \bibinfo{author}{\bibfnamefont{S.~P.} \bibnamefont{Alpay}}, \bibnamefont{and}
  \bibinfo{author}{\bibfnamefont{S.~K.} \bibnamefont{Nayak}},
  \bibinfo{journal}{Phys. Rev. B} \textbf{\bibinfo{volume}{99}},
  \bibinfo{pages}{014101} (\bibinfo{year}{2019}),
  \urlprefix\url{https://link.aps.org/doi/10.1103/PhysRevB.99.014101}.

\bibitem[{\citenamefont{Medvedeva et~al.}(1993)\citenamefont{Medvedeva,
  Turzhevsky, Gubanov, and Freeman}}]{PhysRevB.48.16061}
\bibinfo{author}{\bibfnamefont{N.~I.} \bibnamefont{Medvedeva}},
  \bibinfo{author}{\bibfnamefont{S.~A.} \bibnamefont{Turzhevsky}},
  \bibinfo{author}{\bibfnamefont{V.~A.} \bibnamefont{Gubanov}},
  \bibnamefont{and} \bibinfo{author}{\bibfnamefont{A.~J.}
  \bibnamefont{Freeman}}, \bibinfo{journal}{Phys. Rev. B}
  \textbf{\bibinfo{volume}{48}}, \bibinfo{pages}{16061} (\bibinfo{year}{1993}),
  \urlprefix\url{https://link.aps.org/doi/10.1103/PhysRevB.48.16061}.

\bibitem[{\citenamefont{Ravez and Simon}(1988)}]{ravez88}
\bibinfo{author}{\bibfnamefont{J.}~\bibnamefont{Ravez}} \bibnamefont{and}
  \bibinfo{author}{\bibfnamefont{A.}~\bibnamefont{Simon}},
  \bibinfo{journal}{Ferroelectrics} \textbf{\bibinfo{volume}{81}},
  \bibinfo{pages}{309} (\bibinfo{year}{1988}),
  \eprint{https://doi.org/10.1080/00150198808008870},
  \urlprefix\url{https://doi.org/10.1080/00150198808008870}.

\bibitem[{\citenamefont{Ghosez et~al.}(1999)\citenamefont{Ghosez, Cockayne,
  Waghmare, and Rabe}}]{PhysRevB.60.836}
\bibinfo{author}{\bibfnamefont{P.}~\bibnamefont{Ghosez}},
  \bibinfo{author}{\bibfnamefont{E.}~\bibnamefont{Cockayne}},
  \bibinfo{author}{\bibfnamefont{U.~V.} \bibnamefont{Waghmare}},
  \bibnamefont{and} \bibinfo{author}{\bibfnamefont{K.~M.} \bibnamefont{Rabe}},
  \bibinfo{journal}{Phys. Rev. B} \textbf{\bibinfo{volume}{60}},
  \bibinfo{pages}{836} (\bibinfo{year}{1999}).

\bibitem[{\citenamefont{Ghosez et~al.}(1996)\citenamefont{Ghosez, Gonze, and
  Michenaud}}]{Ghosez_1996}
\bibinfo{author}{\bibfnamefont{P.}~\bibnamefont{Ghosez}},
  \bibinfo{author}{\bibfnamefont{X.}~\bibnamefont{Gonze}}, \bibnamefont{and}
  \bibinfo{author}{\bibfnamefont{J.-P.} \bibnamefont{Michenaud}},
  \bibinfo{journal}{Europhysics Letters ({EPL})} \textbf{\bibinfo{volume}{33}},
  \bibinfo{pages}{713} (\bibinfo{year}{1996}).

\bibitem[{\citenamefont{Amoroso et~al.}(2018)\citenamefont{Amoroso, Cano, and
  Ghosez}}]{PhysRevB.97.174108}
\bibinfo{author}{\bibfnamefont{D.}~\bibnamefont{Amoroso}},
  \bibinfo{author}{\bibfnamefont{A.}~\bibnamefont{Cano}}, \bibnamefont{and}
  \bibinfo{author}{\bibfnamefont{P.}~\bibnamefont{Ghosez}},
  \bibinfo{journal}{Phys. Rev. B} \textbf{\bibinfo{volume}{97}},
  \bibinfo{pages}{174108} (\bibinfo{year}{2018}).

\bibitem[{\citenamefont{Auciello et~al.}(1998)\citenamefont{Auciello, Scott,
  and Ramesh}}]{auciello}
\bibinfo{author}{\bibfnamefont{O.}~\bibnamefont{Auciello}},
  \bibinfo{author}{\bibfnamefont{J.~F.} \bibnamefont{Scott}}, \bibnamefont{and}
  \bibinfo{author}{\bibfnamefont{R.}~\bibnamefont{Ramesh}},
  \bibinfo{journal}{Physics Today} \textbf{\bibinfo{volume}{51}},
  \bibinfo{pages}{22} (\bibinfo{year}{1998}),
  \eprint{https://doi.org/10.1063/1.882324},
  \urlprefix\url{https://doi.org/10.1063/1.882324}.

\bibitem[{\citenamefont{Spaldin}(2012)}]{spaldin2012}
\bibinfo{author}{\bibfnamefont{N.~A.} \bibnamefont{Spaldin}},
  \bibinfo{journal}{Journal of Solid State Chemistry}
  \textbf{\bibinfo{volume}{195}}, \bibinfo{pages}{2} (\bibinfo{year}{2012}),
  ISSN \bibinfo{issn}{0022-4596}, \bibinfo{note}{polar Inorganic Materials:
  Design Strategies and Functional Properties},
  \urlprefix\url{https://www.sciencedirect.com/science/article/pii/S0022459612003234}.

\bibitem[{\citenamefont{Weller et~al.}(2004)\citenamefont{Weller, Hughes,
  Rooke, Knee, and Reading}}]{B401787K}
\bibinfo{author}{\bibfnamefont{M.~T.} \bibnamefont{Weller}},
  \bibinfo{author}{\bibfnamefont{R.~W.} \bibnamefont{Hughes}},
  \bibinfo{author}{\bibfnamefont{J.}~\bibnamefont{Rooke}},
  \bibinfo{author}{\bibfnamefont{C.~S.} \bibnamefont{Knee}}, \bibnamefont{and}
  \bibinfo{author}{\bibfnamefont{J.}~\bibnamefont{Reading}},
  \bibinfo{journal}{Dalton Trans.} pp. \bibinfo{pages}{3032--3041}
  (\bibinfo{year}{2004}), \urlprefix\url{http://dx.doi.org/10.1039/B401787K}.

\bibitem[{\citenamefont{Knee et~al.}(2000)\citenamefont{Knee, Rainford, and
  Weller}}]{B006120O}
\bibinfo{author}{\bibfnamefont{C.~S.} \bibnamefont{Knee}},
  \bibinfo{author}{\bibfnamefont{B.~D.} \bibnamefont{Rainford}},
  \bibnamefont{and} \bibinfo{author}{\bibfnamefont{M.~T.}
  \bibnamefont{Weller}}, \bibinfo{journal}{J. Mater. Chem.}
  \textbf{\bibinfo{volume}{10}}, \bibinfo{pages}{2445} (\bibinfo{year}{2000}),
  \urlprefix\url{http://dx.doi.org/10.1039/B006120O}.

\bibitem[{\citenamefont{Knee and Weller}(2001)}]{B102092G}
\bibinfo{author}{\bibfnamefont{C.~S.} \bibnamefont{Knee}} \bibnamefont{and}
  \bibinfo{author}{\bibfnamefont{M.~T.} \bibnamefont{Weller}},
  \bibinfo{journal}{J. Mater. Chem.} \textbf{\bibinfo{volume}{11}},
  \bibinfo{pages}{2350} (\bibinfo{year}{2001}),
  \urlprefix\url{http://dx.doi.org/10.1039/B102092G}.

\bibitem[{\citenamefont{Hyatt et~al.}(2003)\citenamefont{Hyatt, Hriljac, and
  Comyn}}]{HYATT2003837}
\bibinfo{author}{\bibfnamefont{N.~C.} \bibnamefont{Hyatt}},
  \bibinfo{author}{\bibfnamefont{J.~A.} \bibnamefont{Hriljac}},
  \bibnamefont{and} \bibinfo{author}{\bibfnamefont{T.~P.} \bibnamefont{Comyn}},
  \bibinfo{journal}{Materials Research Bulletin} \textbf{\bibinfo{volume}{38}},
  \bibinfo{pages}{837} (\bibinfo{year}{2003}), ISSN \bibinfo{issn}{0025-5408},
  \urlprefix\url{https://www.sciencedirect.com/science/article/pii/S0025540803000321}.

\end{thebibliography}

\end{document}